# Regional Economic Impacts of the Just Energy Transition: Lessons for Coal Regions

Imke Rhoden[1*], Jae-Hyuck Lee[2]

[1] Forschungszentrum Jülich GmbH, Institute of Climate and Energy Systems, Jülich Systems Analysis, 52425 Jülich, Germany.
[2] Korea Environment Institute, Sejong, South Korea
*Corresponding author: i.rhoden@fz-juelich.de, ORCID: 0000-0001-9533-0536. This work was supported by the Helmholtz Association, Germany, under the program "Energy System Design".

Abstract:
The coal phase-out's regional economic impact is a key challenge of the energy transition, as employment and fiscal dependence in coal regions face structural adjustment without automatic market solutions. Analyzing European Union NUTS 2 regions from 2000–2022 with fixed effects and clustered errors, coal regions show a consistent 1.1 percentage points unemployment premium and grow faster in gross domestic product per capita at 0.2 percentage points annually, indicating a hollowing-out process where population exit raises per-capita output while employment conditions worsen. Spatial analysis shows strong geographic clustering, supporting coordinated local and sectoral targeted transition policies. South Korea's rapid phase-out, with Chungnam as a major coal-power region, underscores the need for proactive national support to enable concrete regional action before plants shut down.

# I. Introduction

The phasing out of coal happens on a global scale, and it is a major structural transition process. Coal has historically been a main fuel in Europe's economy, with annual hard coal consumption stabilizing around 300 million tons from 1999 for nearly a decade. Post-2008 declines and the COVID-19 pandemic led to a sharp drop, with 2024 consumption estimated at 107 million tons, 51% less than in 2018. Brown coal consumption also fell to 198 million tons in 2024, a 47% decrease since 2018. Since 1990, coal's role in electricity production has steadily declined, with minor rebounds in 2011 and 2021-2022. The share of hard coal in electricity generation was surpassed by natural gas in 2005, hydro and wind in 2018, brown coal in 2019, and solar in 2022, with solar overtaking brown coal again in 2023[1].

Korea's accelerating coal phase-out, with 40 plant closures confirmed by 2040 and Chungcheongnam-do Province accounting for approximately half of the national coal generating capacity, provides the motivating regional policy context for this analysis of European transition experience. The regional concentration of coal capacity closely parallels the European pattern. Chungcheongnam-do (Chungnam) Province hosts approximately 18 GW of coal-fired generating capacity, around half of South Korea's total, and accounts for approximately 25% of national greenhouse gas emissions[2].

Usage varies across countries, and its role is diminishing as renewables expand, driven by efforts to reduce greenhouse gases, tighter post-2020 emission standards, high CO2 allowance prices, and possible restrictions on coal in future capacity plans. However, this decline risks negative impacts on employment and regional economies in coal mining and power regions. Early action is needed to develop alternative opportunities to sustain regional jobs and economic growth[3]. The core tension lies in a globally necessary decarbonization that is mostly accounted for on national levels, yet it is regionally disruptive.

In this light, a just transition can be a policy response to this asymmetry. A just transition where climate policy considers equity, ensuring regions and workers are not disproportionately burdened, is essential[4]. Within a broad just transition framework, regional development policies are needed to ensure a dialogue between stakeholders, to account for regional structural differences[5].

This study contributes regional economic evidence from European Union (EU) coal phase-outs, briefly reviews the emerging policy toolkit, and derives transferable lessons with relevance to Korea. The study is structured as follows: Section 2 will introduce the regional economic relevance and structural economic impacts of phasing out coal, and Section 3 will present the European coal regions and the economic importance of coal on the labor market and the economy. Section 4 presents policy responses, and Section 5 provides an empirical analysis of regional economic data for Europe. Section 6 suggests lessons for Korea, and Section 7 concludes.

[1] Eurostat, 2025.
[2] Jeyakumar & Kim, 2018; Powering Past Coal Alliance (PPCA), 2025.
[3] Alves Dias et al., 2018.
[4] UNFCCC, 2015.
[5] Galgóczi, 2020.

# II. Coal Phase-Out and Regional Economic Vulnerability

Coal regions are structurally different from other regions undergoing economic structural change because they contain high sectoral concentration, where a single industry accounts for disproportional shares of regional employment, economic output, and tax revenue. These regional economies are only capable of being resistant or resilient to external shocks to a limited extent, and market-driven recovery is highly unlikely without targeted policy intervention. The sectoral concentration on coal mining or power generation is compounded by spatial lock-ins, where capital, labor, and infrastructure are sector-specific and geographically immobile. When the industry declines, firms and workers cannot easily redeploy to alternative sectors. Mining skills are among the most occupation-specific in the economy, acquired through years of practice in conditions that bear little resemblance to other industries, and the workers who hold them are typically concentrated in communities with few alternative employers within commuting distance. Resulting is a regional economy characterized by limited sectoral diversification, high dependence on non-market income transfers, and geographic peripherality. In contrast to other structural change-regions, distance from agglomeration economies and labor markets hinders quick adjustment[6]. These differences explain why standard policy instruments have consistently underperformed in coal regions and need to be refined to be more effective[7].

The economic consequences of coal phase-out propagate through four distinct transmission channels that together constitute the analytical framework of this study. The first and most visible channel is direct employment loss in mining and power generation. As plants close and mines are exhausted or made economically unviable by carbon pricing, the jobs directly attached to coal operations disappear. The second channel operates through local supply chains and induced demand. Every direct coal job supports additional employment in equipment maintenance, logistics, catering, retail, and professional services, and indirect and induced employment represents approximately 30 to 60 percent of direct employment across EU coal regions, meaning total job losses substantially exceed forecasts from plant closure announcements[8]. The third channel is fiscal base erosion. Energy installations generate property taxes, corporate levies, and local authority revenues that fund public services in coal municipalities, and their closure or relocation withdraws this fiscal base precisely when demand for unemployment support, retraining services, and social care is rising. This constrains the local policy response from the fiscal side. The fourth channel is outmigration and demographic decline, as decreasing employment opportunities lead to younger and more mobile workers leaving first. This erosion of the human capital base reduces local consumer demand, accelerates the decline of retail and service sectors, and produces a self-reinforcing spiral in which population loss generates further economic deterioration that generates further population loss. Each channel amplifies the others, and the interaction between them is what transforms a sectoral employment shock into a structural regional crisis that persists long after the initial plant closures.

[6] Krugman, 1991.
[7] Rodríguez-Pose, 2018.
[8] Alves Dias et al., 2018.

The temporal dimension of the adjustment in coal regions requires policy to distinguish between short-run and long-run structural divergences. In the short run, phasing out coal results in immediate income shocks for displaced workers and communities. Passive income support, social protection, and direct employment services need to prevent adverse economic consequences like poverty and maintain social cohesion in these situations. This challenge becomes a structural issue in the long run, because the regional economy needs to be rebuilt and diversified to attract new companies, retain population, and develop institutional capacity for growth. In contrast to passive support that might stabilize a society in the region, this needs active investment so that they can overcome structural dependence and economic marginalization even one generation after the phase out[9].

A just transition contains these challenges within a normative framework. Within, distributional equity ensures that the costs of decarbonization are not disproportionately burdening workers, communities, and regions most dependent on fossil fuels. Procedural equity provides affected communities with transition programs that include meaningful participation in their design and governance, and not just offer external interventions[10]. In practice, the European experience shows that transition programs designed without genuine regional input have consistently suffered from poor local fit, weak legitimacy, and lower uptake than programs co-designed with affected communities and workers[11]. Generally, transitions like a coal phase out need a comprehensive assessment and targeted policy response as they are highly complex phenomena that include socio-economic, political, and environmental dimensions that are interconnected to power dynamics at multiple regional and sectoral levels[12].

# III. Regional Economic Impacts in Coal Regions

The top 10 most exposed NUTS 2 (Nomenclature of territorial units for statistics) regions span across Poland, Bulgaria, Romania, the Czech Republic, and Germany. Illustrating the economic importance of coal mining in European regions in Table 1, the largest shares of employment are found in Poland, where in Śląskie alone approximately 117,000 direct and indirect coal jobs were located, which is the most prominent exposure case in the EU. Even though the absolute number varies, every direct job in coal carries a multiplier of 0.4-1.2, additionally affecting indirect jobs through local supply chains and induced demand in other industry sectors. Therefore, phasing out coal triggers losses well beyond the coal sector[13].

In history, the transition of the coal sector is linked with a persistent socio-economic structural change. Generally, coal regions face a structural challenge of economic coal lock-in, as dependency in mining regions limited the development of other sectors and resulted in lower gross domestic product (GDP) per capita, like in Poland, Germany, Spain, and the United

[9] Galgóczi, 2018.
[10] European Commission, 2021; Pai et al., 2020.
[11] Stevis & Felli, 2015.
[12] Barragan-Contreras et al., 2025.
[13] Spencer et al., 2018.

Kingdom. Even if coal mining itself is not highly capital-intensive, it is often connected to high labor productivity and a high elasticity of the latter, especially locally[14].

After the 1984-85 UK miners' strike, coalfield communities saw decades of high unemployment due to specialized skills and limited local jobs. Similarly, in Germany's Ruhr region, mine closures led older workers to exit the labor force permanently, while younger workers left the area, weakening the regional skill base over time[15]. Furthermore, falling tax revenues while expenditures rise in affected municipalities further weaken coal-dependent regions . In Germany, the labor force in lignite regions is decreasing due to the phase-out, as a result of migration outflows. Unemployment and lack of opportunities as a result of the structural change are especially prominent in East Germany, as the region is relatively homogeneously structured, even though negative effects also show up in West German coal regions[16].

**Table 1. Top 10 most exposed NUTS 2 regions**

| NUTS 2 | Country | Region Name | Direct Jobs | Indirect Jobs Total | Total Jobs |
|---|---|---|---|---|---|
| PL22 | PL | Śląskie | 82459 | 34536 | 116995 |
| PL11 | PL | Łódzkie | 8926 | 19459 | 28385 |
| BG34 | BG | Yugoiztochen | 12658 | 12063 | 24721 |
| RO41 | RO | Sud-Vest Oltenia | 13139 | 8214 | 21353 |
| CZ04 | CZ | Severozápad | 9731 | 10310 | 20041 |
| CZ08 | CZ | Moravskoslezsko | 10554 | 3840 | 14394 |
| DEA2 | DE | Köln | 5728 | 8275 | 14003 |
| DEA3 | DE | Münster | 10009 | 3365 | 13374 |
| PL41 | PL | Wielkopolskie | 3385 | 8090 | 11475 |
| DEA1 | DE | Düsseldorf | 4594 | 6754 | 11348 |

Note: Selected by total coal employment (direct + indirect).
Source: Alves Dias et al., 2018..

Table 2 illustrates that coal regions averaged 17,415 PPS per capita in 2005 vs. 22,260 for non-coal, which yields a 22% raw deficit. The absolute gap has widened to 5,723 PPS by 2022 (non-coal: 34,752; coal: 29,029), even as the relative gap narrowed from 78% to 84% of non-coal levels. Both things are simultaneously true, and both matter: the catching-up is real, but it has not closed the gap in absolute terms.

The divergence between output indicators and labor market conditions marks regional hollowing-out as a structural transition, not just a statistical artifact. When workers leave the labor force through early retirement, disability, or outmigration, GDP per capita may stay stable or improve as the resident population declines faster than GDP. This is documented in European regions, especially UK coalfields, where unemployment undercounts distress, with displaced workers often on long-term sickness or inactivity rather than registered unemployment. Thus, GDP growth in a coal region doesn't necessarily indicate a successful transition without considering labor market and demographic trends[17]. Additionally, outmigration can create a negative spiral when younger, higher-skilled workers leave first, eroding the human capital base and shrinking demand for local services, which accelerates further economic decline[18].

[14] Alves Dias et al., 2018; Spencer et al., 2018.
[15] Beatty & Fothergill, 1996.
[16] Oei et al., 2020.
[17] Beatty & Fothergill, 2023.

**Table 2. Descriptive statistics**

| Group | Year | N | Mean GDP | Median GDP | Mean Unemployment |
|---|---|---|---|---|---|
| Non-coal | 2005 | 253 | 22260 | 22200 | 8.4 |
| Non-coal | 2010 | 261 | 25077 | 24300 | 8.9 |
| Non-coal | 2015 | 261 | 27129 | 26050 | 9.2 |
| Non-coal | 2019 | 254 | 30547 | 28950 | 6.4 |
| Non-coal | 2022 | 216 | 34752 | 32000 | 6.2 |
| Core coal | 2005 | 38 | 17415 | 17550 | 10.2 |
| Core coal | 2010 | 38 | 20258 | 19100 | 10 |
| Core coal | 2015 | 36 | 22054 | 20650 | 9.5 |
| Core coal | 2019 | 36 | 25200 | 24400 | 5.8 |
| Core coal | 2022 | 28 | 29029 | 29350 | 5.9 |

Note: Mean GDP per capita and unemployment rate for coal vs. non-coal regions at five-year intervals (2005-2022). UK regions are only included from 2000 to 2020 due to Brexit 2020.
Source: Alves Dias et al., 2018; Eurostat, 2026a, 2026c.

Under Korea's latest 2035 Nationally Determined Contribution (NDC), the share of renewables in power generation, which stood at around 9% in 2024, is targeted to reach at least 30% by 2035, while coal-fired power generation is expected to be phased down with a view to achieving a coal phase-out by 2040[19]. After joining the Powering Past Coal Alliance at COP30, South Korea confirmed that 40 of its 61 existing coal-fired units are scheduled for phase-out by 2040, while the treatment of the remaining units requires further public discussion and feasibility review[20]. Chungnam is therefore best understood as a coal-power host region rather than a coal-mining region: it hosts a large share of Korea's coal-fired power fleet and is also one of the country's largest greenhouse-gas-emitting provinces[21]. The relocation of coal-related firms from Chungnam may be interpreted as an early warning sign of supply-chain erosion, although further municipal-level employment, population, and fiscal data are needed to determine whether a broader hollowing-out dynamic is already emerging. Public support for managed transition can coexist with concerns about electricity prices and local adjustment costs. The window for proactive action in Chungnam is open but narrowing; once supply chains weaken and outmigration rises, reversing the trend becomes more difficult and costly.

# IV. Policy Responses

## 1. EU-level Architecture and Labor Market Policies

To comply with the Paris Agreement, OECD countries need to phase out coal power by 2030 at the latest[22]. European governments that decided to proceed with phasing out coal need to confirm plant closure dates and offer plans to manage a just transition consisting of environmental, societal, and economic responsibilities. So far, 23 countries have announced a phase-out. The Just Transition Fund (JTF) is part of the Just Transition Mechanism and was established as a regulation in 2021 to help achieve energy and climate goals[23]. Aside from financial and practical support, employment structures and investments in affected regions are offered, to which the coal regions belong. EU Commission support can help national and regional institutions to initiate and implement the transition[24]. With 17.5 billion euros for 2021-2027, the JTF supports about 100 NUTS 3 regions in 18 countries, focusing on economic diversification, worker retraining, and clean energy in transition-affected areas. Eligibility depends on territorial just transition plans co-created by governments, regions, and social partners, linking funding to national climate commitments. The JTF marks a key innovation by recognizing regional impacts of climate policy and formalizing compensatory transfers in EU climate governance[25].

[18] Oei et al., 2019.
[19] Republic of Korea, 2025.
[20] Powering Past Coal Alliance (PPCA), 2025.
[21] Governors Association of Korea, 2025.
[22] UNFCCC, 2015.

The Act to Reduce and End Coal-Fired Power Generation translates into law the energy-policy recommendations made by the Commission on Growth, Structural Change and Employment, colloquially known as the Coal Commission. Coal-fired power generation is to be phased out gradually, ending completely no later than the end of 2038. The Act supplements the Structural Development Act enacted in 2020[26]. The legislation commits 40 billion euros aligned with the coal phase-out timetable for the four coal regions, Lusatia, Rheinisches Revier, Mitteldeutsches Revier, and Helmstedter Revier. Funds are managed through federal, state, and regional agencies, covering infrastructure, research, public administration, and business development. The program’s design learns from the Ruhr Area’s deindustrialization, providing long-term investment certainty and strong governance to effectively deploy funds[27].

The UK's coal transition began decades before EU policies, with post-1984-85 miners' strike efforts like the Coalfield Regeneration Trust in 1999, aimed at economic and community recovery. In 2024, the last coal-fired power plant was closed[28]. It shows that reactive policies, arriving after job losses, limited the prevention of long-term unemployment and decline. Yet, the Trust's long-lasting community-led approach highlights the importance of continuity and local control in managing structural change[29].

Poland highlights the political economy challenges in Europe's coal transition, as there is no coal phase-out discussion. Śląskie Province employs about 80,000 people in coal mining and retains strong political influence. Despite EU commitments, transition delays persist, with the 2021 coal agreement extending mine operations to 2049. This compresses the adjustment window, leaving a large, exposed workforce and underdeveloped transition institutions. Poland's example shows that political delays defer costs, which then become more expensive and disruptive than an early, managed transition[30].

---

[23] European Commission, 2021.
[24] European Commission, 2019.
[25] European Commission, 2021.
[26] Bundesregierung, 2020.
[27] Bundesministerium für Wirtschaft und Energie (BMWE), 2026.
[28] Beyond Fossil Fuels, 2026.
[29] Beatty & Fothergill, 1996; Trust, 2026.
[30] Beyond Fossil Fuels, 2026.

Regarding labor market policies relevant for coal regions, early retirement has been the most widely deployed labor market instrument in European coal transitions. The German Anpassungsgeld, providing income support to coal workers from age 58 until statutory pension age, is the most documented example[31]. It is politically effective in reducing headline unemployment during rapid mine closure, but costly, age-restricted, and incapable of generating new economic activity or preparing communities for post-coal employment structures. Retraining and reskilling programs have produced rather mixed results. They work mostly for participants under 50, are occupation-specific rather than generic training, and target employers actively involved in program design. However, these conditions are difficult to meet in geographically remote, mono-structural regions where the primary barrier to employment is limited local demand rather than individual skill deficits[32]. The fundamental distinction is between passive instruments, which provide essential income security but do not generate employment, and active instruments. These can address the regional dimension of structural change better by offering, e.g., employer subsidies, place-based investment, sector-specific retraining linked to committed firms[33]. The European evidence consistently shows that passive support alone achieves short-run social stabilization but produces persistent long-run underperformance. Durable recovery requires active instruments deployed from the outset rather than introduced after the labor market has already deteriorated[34].

## 2. Regional Economic Development

Investment subsidies, special economic zones, and infrastructure investment have all featured in European coal transition programs, with different effectiveness. Firm-level subsidies and enterprise zones generate job creation in the short run but frequently attract inefficient investment that exits when incentive periods expire, without building the supply chain linkages or institutional relationships that sustain regional economic capacity[35]. Infrastructure investment has a stronger and more durable evidence base as it reduces effective distance to adjacent economic centers, lowers logistics costs for potential investors, and enables residents to access employment in neighboring regions during the adjustment period[36]. This reduces outmigration pressure. The critical variable across all these instruments, however, is institutional capacity, so regional governments and development agencies are able to design projects, manage procurement, coordinate across governance levels, and leverage private co-investment[37]. The contrast between, e.g., the German Lusatia and Greek Dytiki Makedonia illustrates that both regions carry comparable coal employment exposure relative to regional population, yet Lusatia benefits from decades of institutional infrastructure. There exist established development agencies, strong federal administrative capacity, and embedded private sector relationships, while Dytiki Makedonia operates with limited regional governance capacity and effectively centralized transition planning. One result of that is slower fund absorption, less locally adapted project design, and weaker private investment leverage in the

[31] Bundesministerium für Wirtschaft und Energie (BMWE), 2025.
[32] Galgóczi, 2018.
[33] Pai et al., 2020.
[34] Oei et al., 2020.
[35] Neumark & Simpson, 2015.
[36] Crescenzi & Rodríguez-Pose, 2012.
[37] McCann & Varga, 2018.

Greek case relative to the German[38]. This comparison carries a direct implication for Korea: funding without commensurate investment in the institutional capacity to deploy does not produce transition. Building regional governance capacity is itself a policy investment that must precede or accompany financial transfers.

## 3. Timing

The most important lesson from European coal transitions, however, is not which instrument is most effective, but when support is deployed relative to the phase-out itself. Proactive transition policy initiated before plant closures produces substantially better regional economic outcomes than reactive policy initiated in response to employment loss, operating through multiple channels simultaneously[39]. It provides the planning certainty that private investors require, allows institutional infrastructure to be built before it is urgently needed, preserves community social cohesion, and reduces the long-run fiscal cost of transition by preventing the accumulation of unemployment, health, and social costs that follow unmanaged regional decline[40]. The Ruhr Area underwent a contraction over a fifty-year horizon with active structural policy running in parallel throughout, building the institutional depth that absorbed final closures without acute crisis, even if not completely frictionless and supported by an innovation and education environment[41]. By contrast, Lusatia faced an accelerated post-reunification transition without comparable preparation and required the exceptional intervention of the Structural Development Act to establish the conditions for managed adjustment decades later[42]. Equally important is timeline credibility, because repeated revision of closure schedules, whether by extension, as in Poland, or by acceleration without accompanying support commitments, destroys forward-looking investment decisions. Still, transition policy depends upon these investments to generate leverage[43]. Recent evidence from Germany, where market forces are driving coal exits ahead of the statutory 2038 deadline even amid an active energy crisis, confirms that the economic case for transition management rests on long-run cost trajectories rather than cyclical price signals[44]. For Korea, where 40 plant closures are confirmed by 2040, and the first retirements in Chungnam are already underway, the institutional and fiscal infrastructure for regional transition must be established now[45]. The European evidence is unambiguous and illustrates that regions that managed coal transition most successfully were those that began preparing before it was urgent[46].

[38] Alves Dias et al., 2018.
[39] Reitzenstein et al., 2021.
[40] Galgóczi, 2018.
[41] Hospers, 2004.
[42] Reitzenstein et al., 2021.
[43] Rentschler & Bazilian, 2017.
[44] Wehrmann, 2026.
[45] Gordon et al., 2025.
[46] Oei et al., 2020; Oei et al., 2019.

# V. Quantifying the Coal Region Penalty

## 1. Data and Methodology

To set up the empirical analysis, data about gross domestic product per capita in purchasing power standards and the unemployment rate of people aged 20-64 are obtained for the years 2000-2022, making a panel of 23 years. The regional entity corresponds to the NUTS 2 level. To control for urbanization, population density is used[47]. To classify the coal regions, NUTS 2 level regions with active or recently closed mining operations in 2018 are designated as the "core" treatment group, which results in 38 regions. Regions with power plants but no mining are kept separate as "power-only" regions and are used for robustness checks; these are 67 regions[48]. Regions from the United Kingdom are only included for 2000-2020 due to Brexit. This reduces the sample size by 10 regions after 2020.

As a model specification, a two-way fixed effects (FE) model with year and country FE is used. Because the coal region status is time invariant, region FE cannot be included, so all comparisons are within the same country across coal and non-coal regions. This conservative approach mitigates cross-country level GDP differences. The standard errors (SE) are clustered by NUTS 2 region throughout, for models (M1)-(M6):

$$lnGDP_{it} = \alpha + \beta \cdot Coal_i + \gamma_t + \delta_c + \varepsilon_{it} \quad (1)$$

where ln $GDP_{it}$ is the natural log of GDP per capita (PPS, EU27=100) in region $i$ in year $t$, $\text{Coal}_i$ is a binary indicator equal to 1 for core coal mining regions and 0 otherwise[49], $\gamma_t$ are year FE absorbing common macroeconomic shocks, $\delta_c$ are country FE ensuring all comparisons are made within the same country, and $\varepsilon_{it}$ is the error term clustered by NUTS 2 region. The coefficient $\beta$ is the within-country log GDP per capita differential associated with coal region status, and $(\exp(\hat{\beta}) - 1) \times 100$ gives the approximate percentage difference.

Model M1 omits $\delta_c$ to show the raw cross-country comparison. Models M5 and M6 restrict the sample to 2013-2022 and to non-power-only control regions, respectively, as robustness checks.

For the unemployment rate outcome (Model M4), the dependent variable enters in levels:

$$U_{it} = \alpha + \beta \cdot Coal_i + \gamma_t + \delta_c + \varepsilon_{it} \quad (2)$$

where $U_{it}$ is the unemployment rate (%, ages 20-64) in region $i$ in year $t$. The coefficient $\beta$ is interpreted directly as a percentage point differential.

For the GDP growth rate outcome (Model M3):

$$g_{it} = \alpha + \beta \cdot \text{Coal}_i + \gamma_t + \delta_c + \varepsilon_{it} \quad (3)$$

where $g_{it} = \left(\frac{GDP_{it}}{GDP_{i,t-1}} - 1\right) \times 100$ is the annual GDP per capita growth rate in percentage points.

[47] Eurostat, 2026a, 2026b, 2026c.
[48] Alves Dias et al., 2018.
[49] Alves Dias et al., 2018.

The identification relies on within-country variation between coal and non-coal regions over time. As coal regions do not change, this might reflect unobserved regional characteristics that are correlated with GDP outcomes. This is partially addressed by controlling for initial GDP level and population density, and by restricting comparisons to within-country. As further checks, Conley SE and spatial econometric model specifications are implemented (spatial autoregressive and spatial error models after testing for spatial autocorrelation and dependence)[50]. Results should be interpreted as exploratory and descriptive of the conditional association between coal specialization and economic outcomes rather than as causal estimates.

For the augmented specifications, Models AM2 and AM8 add population density as a control for urbanization and remoteness:

$$\ln GDP_{it} = \alpha + \beta \cdot \text{Coal}_i + \theta \cdot \ln D_{it} + \gamma_t + \delta_c + \varepsilon_{it} \quad (4)$$

where $D_{it}$ is population density Eurostat, 2026b..

Model AM4 adds initial GDP per capita as a convergence control:

$$g_{it} = \alpha + \beta \cdot \text{Coal}_i + \phi \cdot \ln GDP_{i,2000} + \gamma_t + \delta_c + \varepsilon_{it} \quad (5)$$

where $\ln GDP_{i,2000}$ is the log GDP per capita in the base year 2000. A negative $\hat{\phi}$ would indicate conditional convergence, poorer regions growing faster, all else equal. The stability of $\hat{\beta}$ across Models AM3 and AM4 confirms that the growth premium is not driven by convergence dynamics.

## 2. Descriptive Evidence

For GDP per capita, all NUTS 2 regions in Europe are investigated. Regional GDP per capita in PPS is used throughout as the standard measure of regional economic output. Results are qualitatively identical when estimated using nominal EUR per capita values. To illustrate the within-country deviation, the country's mean is subtracted from each region's value in each year in Figure 1. This removes the country-level income differences and aligns the visual evidence with the regression specification. The zero line shows the country's average. The coal regions are consistently below their national mean in GDP per capita throughout 2000-2022. The gap is persistent and does even increase over time, up to a value of -2,500 GDP per capita less than other regions, despite two decades of cohesion policy in the EU. There is considerable variation in the data. However, most coal regions stay below the national average, and even farther below the non-coal regions, which, on average, are slightly above the national mean.

[50] Conley, 1999.

**Figure 1. GDP per capita: Deviation from country mean, coal vs. non-coal NUTS 2 regions**

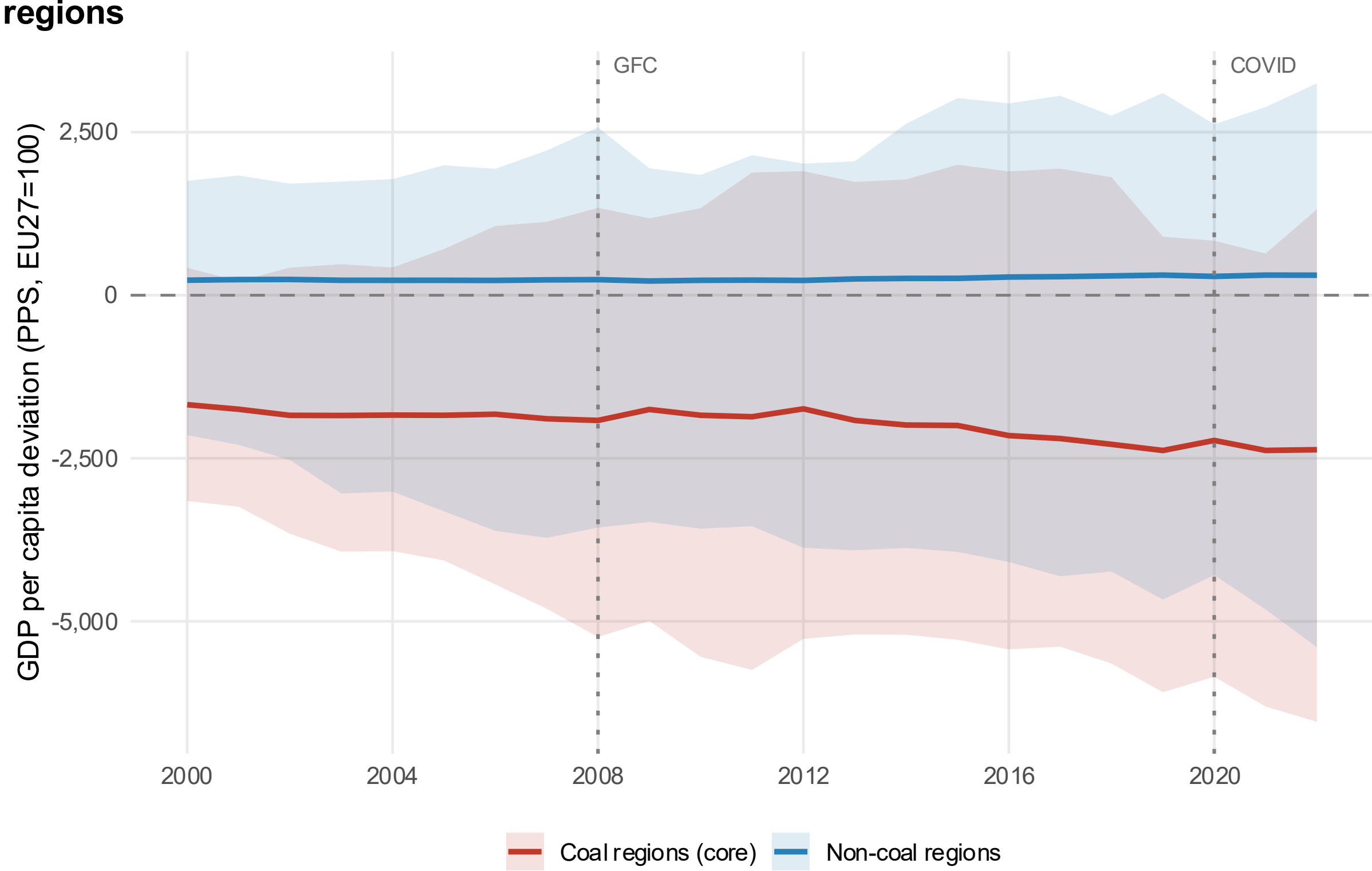

Note: 2000-2022, mean ± Interquartile Range, zero = country average. Deviation = region minus own-country mean in each year. GFC: Global Financial Crisis, COVID: COVID-19 pandemic. pp: percentage points.
Source: Eurostat, 2026a.

Regarding unemployment illustrated in Figure 2, while all regions exhibit fluctuations of the unemployment rate, coal regions carry a persistent unemployment premium relative to their national mean. The post-2019 convergence in the data shows a composition artifact and not genuine improvement, as the regions in the UK drop out of the dataset. Figure 2 again shows the considerable difference between coal regions and non-coal regions, which exhibit, on average, lower unemployment rates of about 2 percentage points, even though there seems to be a convergence pattern decreasing the gap slightly.

**Figure 2. Unemployment Rate: Deviation from country mean, coal vs. non-coal NUTS 2 regions**

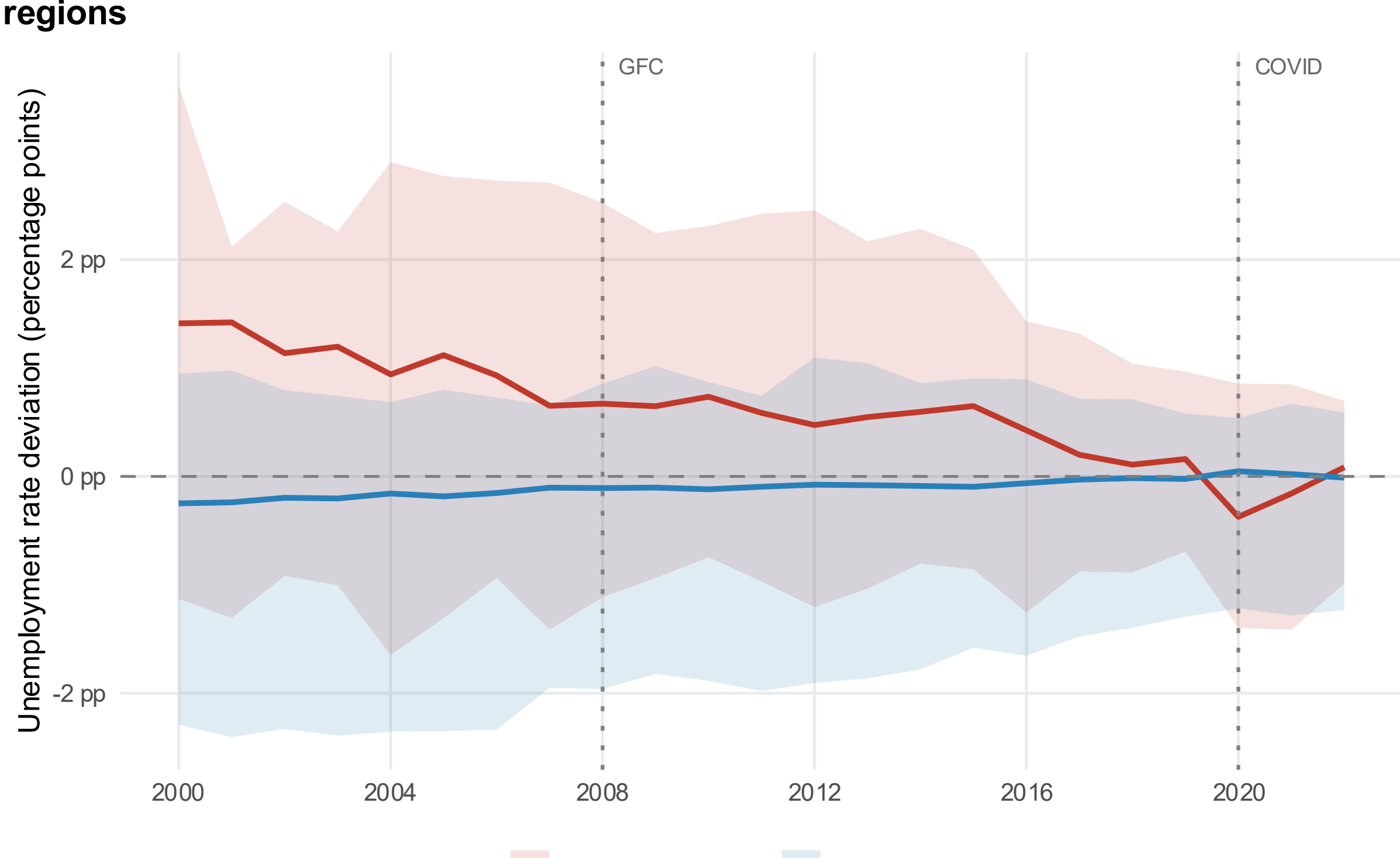

Note: 2000-2022, ages 20-64, mean ± Interquartile Range, zero = country average. Deviation = region minus own-country mean in each year. GFC: Global Financial Crisis, COVID: COVID-19 pandemic. pp: percentage points.
Source: Eurostat, 2026c.

## 3. Regression Results

The regression results confirm the persisting unemployment premium of the coal regions. Table 3 illustrates the baseline panel regression results. Model M1 shows raw cross-country comparisons without country FE and coal regions exhibiting a GDP per capita roughly 19.2% lower than the EU average. With country FE in Model M2, this drops to 6.0% and loses statistical significance. The large drop itself suggests that the majority of the coal penalty in coal regions reflects only cross-country heterogeneity. Model M3 turns to GPD per capita growth as the dependent variable. It shows that coal regions grow significantly faster than non-coal regions in the same countries, with a premium of 0.23 percentage points per year. Faster GDP growth in the presence of a persistent unemployment situation, as documented in Model M4, defines this as a hollowing-out dynamic. When workers exit through early retirement, outmigration, or labor force withdrawal, output per remaining capita rises even if total regional output stagnates or even declines. The GDP per capita measure improves on paper while underlying labor market conditions deteriorate. This divergence between GDP and labor market developments is structurally identical to post-industrial European regions following deindustrialization. Furthermore, it shows that transition success has to be measured and reported carefully[51].

[51] Beatty & Fothergill, 2023.

**Table 3. Regional GDP and Unemployment Penalty for Coal Regions**

| | (M1) log GDP | (M2) log GDP +country | (M3) GDP growth | (M4) Unemployment rate | (M5) Post-2013 | (M6) Clean control |
|---|---|---|---|---|---|---|
| Coal region (=1) | -0.192*** | -0.062 | 0.230** | 0.870* | -0.054 | -0.075 |
| | (0.063) | (0.046) | (0.091) | (0.477) | (0.047) | (0.051) |
| Year FE | Yes | Yes | Yes | Yes | Yes | Yes |
| Country FE | No | Yes | Yes | Yes | Yes | Yes |
| Period | 2000-22 | 2000-22 | 2001-22 | 2000-22 | 2013-22 | 2000-22 |
| Control group | All | All | All | All | All | No power-only |
| N | 5485 | 5485 | 5233 | 6012 | 2439 | 4273 |
| $R^2$ | 0.176 | 0.631 | 0.498 | 0.420 | 0.560 | 0.577 |

$p < 0.1$, ** $p < 0.05$, *** $p < 0.01$

Note: Clustered SE by NUTS 2 region. GDP in PPS (EU27=100). Growth in annual percentage points. Coal region = core mining regions.
Source: Alves Dias et al., 2018.

Model M4 estimates the unemployment differential. Coal regions carry a persistent within-country unemployment premium of 0.87 percentage points. While this might be marginal in statistical terms, the direction and magnitude are consistent across all robustness specifications in Table 4, where the estimate increases to 1.06 percentage points after controlling for population density (Model AM8). The unemployment result is therefore the most robust of the three outcomes and carries the strongest policy implication. It shows the labor market consequences of coal region status persist independently of country-level economic conditions and regional urbanization.

Table 4 presents robustness checks. Models (AM2), (AM5), and (AM7) contain heteroscedasticity- and autocorrelation-consistent Conley SE with a 500km spherical distance cutoff[52], correcting for spatial dependence identified by Moran's I[53]. Point estimates are identical to the baseline, even though the SE widen modestly, but all qualitative conclusions are unchanged. The GDP growth premium remains significant after controlling for initial GDP per capita (models AM4 and AM5), confirming the result is not driven by conditional convergence. The unemployment premium strengthens marginally after controlling for population density (AM8), consistent with denser regions having structurally higher urban unemployment independent of coal status.

[52] Conley, 1999.
[53] Moran, 1950.

**Table 4. Robustness: Conley Spatial SE and Augmented Controls**

| | (AM1) GDP Clustered | (AM2) GDP Conley | (AM3) GDP +Population Density | (AM4) Growth Clustered | (AM5) Growth Conley | (AM6) Unemployment Clustered | (AM7) Unemployment Conley | (AM8) Unemployment + Population Density |
|---|---|---|---|---|---|---|---|---|
| Coal region (=1) | -0.062 | -0.069 | -0.032 | 0.214** | 0.192 | 0.870* | 1.061* | 1.061** |
| | (0.046) | (0.059) | (0.041) | (0.092) | (0.117) | (0.477) | (0.563) | (0.499) |
| Log population density | | | 0.116*** | | | | | 0.439 |
| | | | (0.024) | | | | | (0.285) |
| Year FE | Yes | Yes | Yes | Yes | Yes | Yes | Yes | Yes |
| Country FE | Yes | Yes | Yes | Yes | Yes | Yes | Yes | Yes |
| SE type | Clustered | Conley 500km | Clustered | Clustered | Conley 500km | Clustered | Conley 500km | Clustered |
| Population density control | No | No | Yes | No | No | No | No | Yes |
| Initial GDP (convergence control) | No | No | No | Yes | Yes | No | No | No |
| Log GDP per capita (2000) | | | | -0.170 | -0.136 | | | |
| | | | | (0.141) | (0.107) | | | |
| N | 5485 | 5039 | 5039 | 5016 | 4598 | 6012 | 5656 | 5699 |
| $R^2$ | 0.631 | 0.654 | 0.691 | 0.498 | 0.492 | 0.420 | 0.439 | 0.434 |

p < 0.1, ** p < 0.05, *** p < 0.01

Note: Clustered SE by NUTS 2 region, initial GDP = log GDP per capita in 2000. Columns (AM1)(AM4)(AM6): clustered SE by NUTS 2. Columns. (AM2)(AM5)(AM7): Conley (1999) HAC SE, 500km cutoff. Columns (AM3)(AM8): adds log population density. Columns. (AM4)(AM5): adds log GDP per capita 2000 as convergence control. Coal region = core mining regions. GDP in PPS (EU27=100). Source: Own calculation.

**Figure 3. The Coal Region GDP Penalty Over Time**

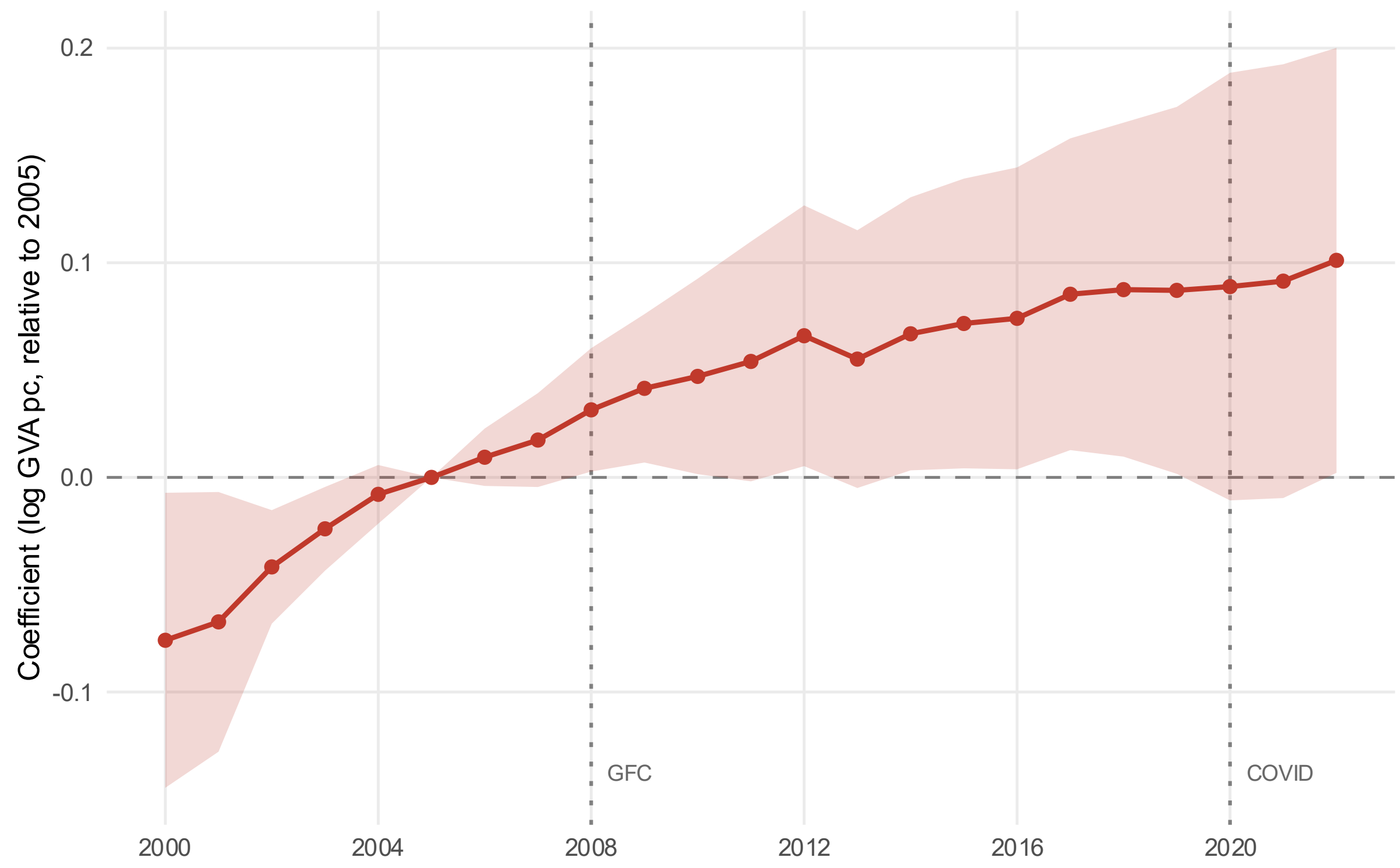

Note: Interaction: coal × year, reference = 2005, 95% confidence interval, region + year FE, SE clustered by NUTS 2 regions, coal × year interaction coefficients, 2000-2022. Negative values = coal regions fell further behind relative to the reference year.
Source: Own calculation.

Figure 3 illustrates the same narrative as the regressions. Coal regions increased compared to non-coal regions in GDP after 2005, peaking around 2015-2018. However, unemployment remained elevated throughout, and the divergence between GDP and labor market trajectories confirms the hollowing-out of coal regions and frames persistent unemployment as regional disadvantage.

# VI. Policy Implications for Korea

## 1. Structural Parallels and Key Differences

Chungnam Province is functionally comparable to European coal transition regions, but the nature of exposure differs. Whereas the European empirical analysis in this paper focuses mainly on coal-mining regions, Chungnam's exposure is concentrated in coal-fired power generation, host municipalities, and related maintenance, logistics, and industrial supply chains. Its regional challenge therefore lies less in mining-sector decline than in the interaction between plant closures, local fiscal dependence, firm relocation, and employment risks in coal-power communities. This makes Chungnam a relevant Korean case for applying the European lessons, while requiring caution against treating it as a direct equivalent of mining-intensive regions such as Lusatia. Firm relocation away from coal-power host areas can constitute early supply-chain erosion that may precede labor-market and demographic decline[54]. European evidence therefore provides a warning for Korea, but the emergence of a broader hollowing-out dynamic should be verified with municipal-level data rather than assumed.

[54] Governors Association of Korea, 2025; Hong, 2014.

Two structural differences are particularly important when transferring European lessons to Korea. First, Korea's coal transition is occurring within a more compressed policy window than the managed decades-long transitions in Germany's Ruhr or even the more accelerated Lusatia case, leaving less time to build institutional capacity, attract replacement investment, and retrain workers before local employment bases contract. Second, Korea already has a legal basis for just transition, including the designation of special districts for just transition under the Framework Act on Carbon Neutrality and Green Growth, but implementation authority, fiscal capacity, and project-delivery capacity at the provincial and municipal levels remain decisive[55]. Germany's coal transitions benefited from strong federal administrative structures, established regional development agencies, and embedded relationships between regional governments, federal ministries, and private investors. Korea's provincial governments have more limited authority over electricity-system planning and industrial siting decisions, which means that transition funds could be underdeployed without parallel investment in regional implementation capacity. These differences point in the same direction: proactive national-level commitment is more urgent in the Korean case, not less.

## 2. Recommendations

The most important lesson from Europe is that transition support must precede phase-out, not follow it. For Korea, the priority is not only to create new instruments, but to operationalize and scale up the existing just-transition framework into a dedicated, multi-year financing mechanism for coal-power host municipalities, co-governed with provincial and local authorities[56]. Chungnam has already begun to develop local just-transition responses, including a Just Energy Transition Fund and proposals for special support for areas affected by coal-fired power plant closures, but these initiatives require national-level scaling, predictable financing, and project-delivery capacity[57]. Moreover, income replacement alone stabilizes households but does not rebuild regional economies, as the European evidence consistently shows. Retraining must be linked to specific absorbing sectors that offer the strongest skill overlap with coal-sector employment profiles. Program design should involve committed employers from the outset rather than delivering generic upskilling into a labor market without absorbing demand. Additionally, deployment capacity is as important as funding volume. The contrast between Lusatia and Dytiki Makedonia shows that regions with stronger administrative institutions absorb transition resources more effectively and leverage greater private co-investment. Korea should establish or strengthen regional transition agencies in affected provinces, simplify procurement rules for transition-designated areas, and provide technical assistance for project design and fund management before closure schedules accelerate. Furthermore, Korea's policy trade-off should not be framed around the use of domestic coal, because Korea is structurally import-dependent in coal. The relevant distinction is between short-run power-system and price concerns and long-run regional adjustment costs. These operate on different time horizons and should not be conflated in policy deliberation. Evidence from Germany, where market forces are driving coal exits ahead of the statutory deadline even amid recent energy price volatility, confirms that the structural economics of coal are deteriorating independently of cyclical price movements. Delay does not reduce transition costs, but it defers and compounds them.

[55] Republic of Korea, 2022.
[56] Republic of Korea, 2022.
[57] Chungcheongnam-do, 2024.

Finally, evidence-based targeting transition funds requires data on regional employment, output, population, firm relocation, and fiscal conditions at the sub-provincial level. The analysis in this paper is constrained by the absence of directly comparable data for Korean coal-power host municipalities. A standardized regional monitoring system for plant-host municipalities such as Boryeong, Dangjin, Seocheon, and Taean should be established immediately, both to inform the design of transition programs and to enable rigorous evaluation of their effectiveness over time.

# VII. Limitations

Several limitations of the empirical analysis should be recognized. The identification strategy depends on cross-sectional differences between coal and non-coal regions within the same country. Since coal region status is fixed over time, it cannot include region FE. Consequently, the estimates reflect the conditional relationship between coal specialization and regional economic outcomes, rather than establishing a causal link to the phase-out policy. They may also partly mirror time-invariant regional traits such as remoteness, historical industrial makeup, or institutional quality, which could be correlated with both coal status and economic performance. Furthermore, the analysis uses data at the NUTS 2 level, which aggregates diverse sub-regions and might mask more pronounced local impacts at the level of individual municipalities or counties directly hosting coal facilities. Moreover, the coal-region classification captures coal employment conditions at a single point in time and doesn't account for variations in phase-out intensity or speed across regions or over the period studied. This treatment groups regions with different transition paths as if they were similar. As the panel only covers EU regions, applying these findings to the Korean context, which varies in governance, regional scale, and transition pace, should be done cautiously. Finally, limited data availability restricted the set of control variables. For instance, the lack of regional sectoral employment shares at NUTS 2 level limited the ability to characterize regional economic structures fully, and excluding UK regions after 2020 results in a composition shift in the post-Brexit period, influencing comparisons between the European regions.

# VIII. Conclusion

The empirical findings show coal regions in Europe exhibit a persistent within-country unemployment premium of approximately 1.1 percentage points relative to non-coal regions in the same country, controlling for population density. Despite this labor market disadvantage, coal regions grow faster in GDP per capita at 0.2 percentage points per year, even after controlling for initial income levels. This aligns with a hollowing-out dynamic, where population and labor force exit increase per-capita output even as employment conditions worsen; rather than indicating a true structural shift, as shown by the stability of the growth coefficient after accounting for convergence. Spatial analysis confirms strong geographic clustering of coal region outcomes, reinforcing the case for coordinated cluster-level transition policy rather than region-by-region intervention.

Taken together, the European experience shows that transition costs are manageable with early, well-funded, and active policy incorporating regional stakeholders. Unmanaged phase-out, however, produces a persistent regional disadvantage that is expensive and slow to reverse. The central lesson is one of timing, with proactive transition support initiated before plant closures. This consistently produces better regional outcomes than reactive policy, which can directly and immediately influence measures to manage, supporting Korea's accelerating phase-out schedule.

In South Korea, the window for proactive policies with the correct timing is open but narrowing. Targeting companies in the Chungnam area is essential to stabilize local labor markets and create new jobs[58]. However, future research should focus on collecting better data and involve regional stakeholders in investigating their preferences and values, and investigating the regional development processes closely[59]. Regional data and support allow improved targeting of funds, which corresponds to the effective implementation of projects.

[58] Baccianti et al., 2022.

[59] Ha & Byrne, 2019.